\documentclass[sigplan,10pt]{acmart}

\setcopyright{none} % No copyright notice required for submissions
\acmYear{2019}
\renewcommand\footnotetextcopyrightpermission[1]{}

\usepackage[normalem]{ulem}

\usepackage{subfig}
\usepackage{fancyvrb}
\usepackage{amsmath}
\usepackage{amssymb}
\usepackage{algorithm}
\usepackage{fullwidth}
\usepackage{algpseudocode}
\usepackage{color}
\usepackage{graphicx}
\usepackage{xspace}
\usepackage{booktabs}
\usepackage{url}
\usepackage{placeins}
\usepackage{etoolbox}
\usepackage{mfirstuc}
\usepackage{ae,aecompl}
\usepackage{bookmark}
\usepackage[font=small]{caption}
\usepackage{paralist}
\usepackage{enumitem}
\usepackage{multirow}
\usepackage{tikz}
\usetikzlibrary{fpu}
\usetikzlibrary{arrows,shapes.geometric,positioning}

\usepackage{fancyvrb}
\usepackage{rotating}
\usepackage{pdflscape}

\usepackage{ifthen}

\newcommand{\DONE}[1]{}
\newcommand{\COMMENT}[1]{}

\newcommand{\figref}[1]{Fig.~\ref{Fi:#1}}

\newcommand{\tabref}[1]{Table~\ref{Ta:#1}}
\newcommand{\secref}[1]{Section~\ref{Se:#1}}

\newcommand{\appref}[1]{Appendix~\ref{Se:#1}}

\newcommand{\figlabel}[1]{\label{Fi:#1}}

\newcommand{\tablabel}[1]{\label{Ta:#1}}
\newcommand{\seclabel}[1]{\label{Se:#1}}

\newcommand{\applabel}[1]{\label{Se:#1}}

\newcommand{\ignore}[1]{}

\newcounter{programlinenumber}

\newboolean{TR}
\setboolean{TR}{false}
\ifthenelse{\boolean{TR}}{

\newcommand{\TrOnly}[1]{#1}
\newcommand{\SubOnly}[1]{}
\newcommand{\TrOnlyInFootnote}[1]{#1}
\newcommand{\TrOnlyInTable}[1]{#1}}
{

\newcommand{\TrOnly}[1]{}
\newcommand{\SubOnly}[1]{#1}
\newcommand{\TrOnlyInFootnote}[1]{}
\newcommand{\TrOnlyInTable}[1]{}}

\newcommand{\sectionette}[1]{\textbf{\emph{#1}}}

\newcommand{\hiddentext}[1]{}

\newcommand{\para}[1]{\vspace{3pt}\noindent\textbf{\textit{#1}} ~~}

\newcommand{\scode}[1]{{\small \texttt{#1}}}
\newcommand{\sname}[1]{{\small \textsc{#1}}}

\renewcommand{\phi}{\varphi}

\newcommand{\tool}{\textsc{TraFix}}

\makeatletter
\newbox\sf@box
\newenvironment{SubFloat}[2][]%
{\def\sf@one{#1}%
\def\sf@two{#2}%
\setbox\sf@box\hbox
\bgroup}%
{ \egroup
\ifx\@empty\sf@two\@empty\relax
\def\sf@two{\@empty}
\fi
\ifx\@empty\sf@one\@empty\relax
\subfloat[\sf@two]{\box\sf@box}%
\else
\subfloat[\sf@one][\sf@two]{\box\sf@box}%
\fi}
\makeatother

\begin{document}

\pagestyle{plain}

\title{Towards Neural Decompilation}

\author{Omer Katz}
\affiliation{
	\institution{Technion}
	\country{Israel}
}
\email{omerkatz@cs.technion.ac.il}

\author{Yuval Olshaker}
\affiliation{
	\institution{Technion}
	\country{Israel}
}
\email{olshaker@cs.technion.ac.il}

\author{Yoav Goldberg}
\affiliation{
	\institution{Bar Ilan University}
	\country{Israel}
}
\email{yoav.goldberg@gmail.com}

\author{Eran Yahav}
\affiliation{
	\institution{Technion}
	\country{Israel}
}
\email{yahave@cs.technion.ac.il}

\date{}
%!TEX root = ./main.tex

\begin{abstract}

We address the problem of automatic \emph{decompilation}, converting a program in low-level representation back to a higher-level human-readable programming language. The problem of decompilation is extremely important for security researchers. Finding vulnerabilities and understanding how malware operates is much easier when done over source code.

The importance of decompilation has motivated the construction of hand-crafted rule-based decompilers. Such decompilers have been designed by experts to detect specific control-flow structures and idioms in low-level code and lift them to source level. The cost of supporting additional languages or new language features in these models is very high. 

We present a novel approach to decompilation based on \emph{neural machine translation}. The main idea is to \emph{automatically learn a decompiler from a given compiler}. Given a compiler from a source language $S$ to a target language $T$, our approach \emph{automatically trains} a decompiler that can translate (decompile) $T$ back to $S$. We used our framework to decompile both \emph{LLVM} IR and \emph{x86} assembly to \scode{C} code with high success rates. Using our \emph{LLVM} and \emph{x86} instantiations, we were able to successfully decompile over 97\% and 88\% of our benchmarks respectively.
\end{abstract}
\maketitle

%%%%%%%%%%%%%%%%%%%%%%%%%%%%%%%%%%%%%%%%%%%%%

\section{Introduction}\seclabel{Intro}

\begin{figure}[tb]
\centering
\includegraphics[width=\columnwidth]{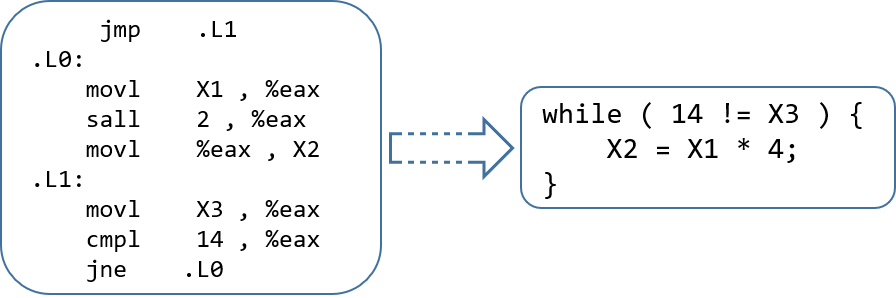}
\begin{tabular}{ccc}(a) & \hspace{1.5in} & (b)\end{tabular}
\caption{Example input (a) and output (b) of decompilation.}
\figlabel{DecompExample}
\end{figure}

Given a low-level program in binary form or in some intermediate representation, \emph{decompilation} is the task of lifting that program to human-readable high-level source code.

\figref{DecompExample} provides a high-level example of decompilation. The input to the decompilation task is a low-level code snippet, such as the one in \figref{DecompExample}(a). The goal of Decompilation is to generate a corresponding equivalent high-level code. The \scode{C} code snippet of \figref{DecompExample}(b) is the desired output for \figref{DecompExample}(a).

There are many uses for decompilation. The most common is for security purposes. Searching for software vulnerabilities and analyzing malware both start with understanding the low-level code comprising the program. Currently this is done manually by reverse engineering the program. Reverse engineering is a slow and tedious process by which a specialist tries to understand what a program does and how it does it. Decompilation can greatly improve this process by translating the binary code to a more readable higher-level code.

Decompilation has many applications beyond security. For example, porting a program to a new hardware architecture or operating system is easier when source code is available and can be compiled to the new environment. Decompilation also opens the door to application of source-level analysis and optimization tools.

\para{Existing Decompilers}
Existing decompilers, such as Hex-Rays~\cite{IDA} and Phoenix~\cite{Phoenix2013}, rely on pattern matching to identify the high-level control-flow structure in a program. These decompilers try to match segments of a program's control-flow graph (CFG) to some patterns known to originate from certain control-flow structures (e.g. \scode{if-then-else} or loops). This approach often fails when faced with non-trivial code, and uses \scode{goto} statements to emulate the control-flow of the binary code. The resulting code is often low-level, and is really assembly transliterated into C (e.g. assigning variables to temporary values/registers, using \scode{goto}s, and using low-level operations rather than high-level constructs provided by the language). While it is usually semantically equivalent to the original binary code, it is hard to read, and in some cases less efficient, prohibiting recompilation of the decompiled code. 

There are \scode{goto}-free decompilers, such as \sname{DREAM++} \cite{Dream2015,Dream++2016}, that can decompile code without resorting to using \scode{goto}s in the generated code. However, all existing decompilers, even \scode{goto}-free ones, are based on hand-crafted rules designed by experts, making decompiler development slow and costly. 

Even if a decompiler from a low-level language $L_{low}$ to a high-level language $L_{high}$ exists, given a new language $L'_{high}$, it is nontrivial to create a decompiler from $L_{low}$ to $L'_{high}$ based on the existing decompiler. There is no guarantee that any of the existing rules can be reused for the new decompiler. 

\para{Neural Machine Translation}
Recent years have seen tremendous progress in Neural Machine Translation (NMT)~\cite{KalchbrennerB13, SutskeverVL14, ChoMBB14}. NMT systems use neural networks to translate a text from one language to another, and are widely used on natural languages. Intuitively, one can think of NMT as encoding an input text on one side and decoding it to the output language on the other side (see \secref{Background} for more details).
Recent work suggests that neural networks are also effective in summarizing source code~\cite{Srinivasan2016,Allamanis2016,Maddison2014,Allamanis2015A,Hu2017,Loyola2017,Allamanis2015B,Amodio2017}. 

Recently, Katz et al.~\cite{saner2018} suggested using neural networks, specifically RNNs, for decompilation.
Their approach trains a model for translating binary code directly to C source code. However, they did not compensate for the differences between natural languages and programming languages, thus leading to poor results. For example, the code they generate often cannot be compiled or is not equivalent to the original source code. Their work, however, did highlight the viability of using Neural Machine Translation for decompilation, thus supporting the direction we are pursuing.
\secref{Related} provides additional discussion of~\cite{saner2018}.

\paragraph{Our Approach}
We present a novel automatic neural decompilation technique, using a two-phased approach. In the first phase, we generate a templated code snippet which is structurally equivalent to the input. The code template determines the computation structure without assignment of variables and numerical constants. Then, in the second phase, we fill the template with values to get the final decompiled program. The second phase is described in~\secref{fixing}.

Our approach can facilitate the creation of a decompiler from $L_{low}$ to $L_{high}$ from every pair of languages for which a compiler from $L_{high}$ to $L_{low}$ exists.

The technique suggested by~\cite{saner2018} attempted to apply NMT to binary code as-is, i.e. without any additional steps and techniques to support the translation. 
We recognize that for a \emph{trainable decompiler}, and specifically an NMT-based decompiler, to be useful in practice, we need to augment it with programming-languages knowledge (i.e. domain-knowledge). Using domain-knowledge we can make translations simpler and overcome many shortcomings of the NMT model. This insight is implemented in our approach as our canonicalization step (\secref{canonization}, for simplifying translations) and template filling (\secref{fixing}, for overcoming NMT shortcomings).

Our technique is still modest in its abilities, but presents a significant step forward towards trainable decompilers and in the application of NMT to the problem of decompilation. The first phase of our approach borrows techniques from natural language processing (NLP) and applies them to programming languages. We use an existing NMT system to translate a program in a lower-level language to a templated program in a higher-level language.

Since we are working on programming languages rather than natural languages, we can overcome some major pitfalls for traditional NMT systems, such as training data generation~(\secref{gen}) and verification of translation correctness~(\secref{eval}).
We incorporate these insights to create \emph{a decompilation technique capable of self-improvement by identifying decompilation failures as they occur, and triggering further training as needed to overcome such failures.} 

By using NMT techniques as the core of our decompiler's first phase, we avoid the manual work required in traditional decompilers. The core of our technique is language-agnostic requiring only minimal manual intervention (i.e., implementing a compiler interface).

One of the reasons that NMT works well in our setting is the fact that, compared to natural language, code has a more repetitive structure and a significantly smaller vocabulary. This enables training with significantly fewer examples than what is typically required for NLP~\cite{nmtChallenges} (See~\secref{Eval}).

\para{Mission Statement}
Our goal is to decompile short snippets of low-level code to equivalent high-level snippets. We aim to handle multiple languages (e.g. \emph{x86} assembly and \emph{LLVM} IR). We focus on code compiled using existing off-the-shelf compilers (e.g. gcc~\cite{gcc} and clang~\cite{clang}), with \emph{compiler optimizations enabled}, for the purpose of finding bugs and vulnerabilities in benign software. More specifically, we do not attempt to handle hand-crafted assembly as is often found in malware.

Many previous works aimed to use decompilation as a mean of understanding the low-level code, and thus focused mostly on code readability. In addition to readability, we place a great emphasis on generating code that is \emph{correct} (i.e., can be compiled without further modifications) and \emph{equivalent to the given input}.

We wish to further emphasize that the goal of our work is not to outperform existing decomopilers (e.g., Hex-Rays~\cite{IDA}). Many years of development have been invested in such decompilers, resulting in mature and well-tested (though not yet perfect) tools.
Rather, we wish to shed light on \emph{trainable decompilation}, and NMT-based decompilation in particular, as a promising alternative approach to traditional decompilation. 
%At its current state, existing decompilers should be to handle benchmarks as the ones we use in our evaluation. 
This new approach holds the advantage over existing decompilers not in its current results, but in its potential to handle new languages, features, compilers, and architectures with minimal manual intervention. We believe this ability will play a vital role as decompilation will become more widely used for finding vulenrabilities. 

\para{Main Contributions}
The paper makes the following contributions:
\begin{itemize}
\item A significant step towards neural decompilation by combining ideas from neural machine translation (NMT) and program analysis. Our work brings this promising approach to decompilation closer to being practically useful and viable.
\item A decompilation framework that automatically generates training data and checks the correctness of translation using a verifier.
\item A decompilation technique that is applicable to many pairs of source and target languages and is mostly independent of the actual low-level source and high-level target languages used.
\item An implementation of our technique in a framework called {\tool} (short for TRAnslate and FIX) that, given a compiler from $L_{high}$ to $L_{low}$ automatically learns a decompiler from $L_{low}$ to $L_{high}$.
\item An instantiation of our framework for decompilation of \scode{C} source code from \emph{LLVM} intermediate representation (IR)~\cite{LLVM} and \emph{x86} assembly. We used these instances to evaluate our technique on decompilation of small simple code snippets.
\item An evaluation showing that our framework decompiles statements in both \emph{LLVM} IR and \emph{x86} assembly back to \scode{C} source code with high success rates.
The evaluation demonstrates the framework's ability to successfully self-advance as needed.
\end{itemize}

\section{Overview}\seclabel{Overview}

In this section we provide an informal overview of our approach.

%% What before how 
\subsection{Motivating Example} 

Consider the x86 assembly example of~\figref{DecompExample}(a). \figref{x86Example} shows the major steps we take for decompiling that example.

\begin{figure*}[tb]
\centering
\includegraphics[width=\textwidth]{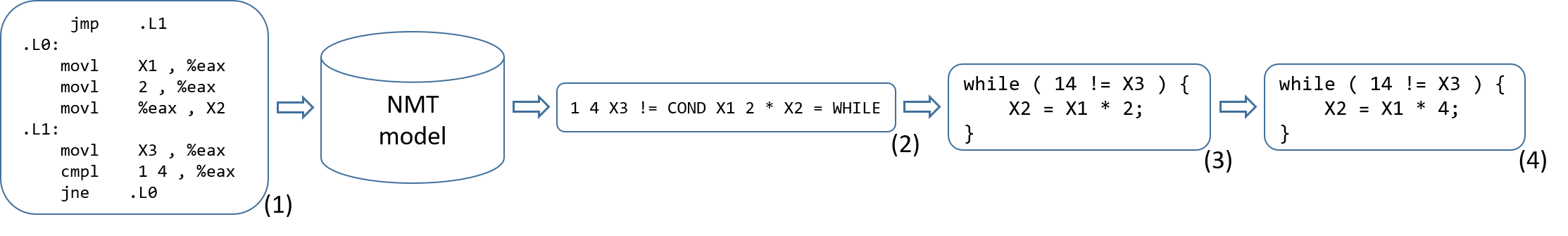}
\caption{Steps for decompiling \emph{x86} assembly to \scode{C}: (1) canonicalized \emph{x86} input, (2) NMT output, (3) templated output, (4) final fixed output.}
\figlabel{x86Example}
\end{figure*}

%\figref{x86Example} presents an example of applying our technique to a sample of \emph{x86} assembly.

The first step in decompiling a given input is applying \emph{canonicalization}. In this example, for the sake of simplicity, we limited canonicalization to only splitting numbers to digits~(\secref{num_abstract}), thus replacing \scode{14} with \scode{1 4}, resulting in the code in block (1). This code is provided to the decompiler for translation.

The output of our decompiler's NMT model is a canonicalized version of \scode{C}, as seen in block (2). In this example, output canonicalization consists of splitting numbers to digits, same as was applied to the input, and printing the code in \emph{post-order}~(\secref{postorder}), i.e. each operator appears after its operands. We apply \emph{un-canonicalization} to the output, which converts it from \emph{post-order} to \emph{in-order}, resulting in the code in block (3). The output of \emph{un-canonicalization} might contain decompilation errors, thus we treat it as a code template. Finally, by comparing the code in block (3) with the original input in~\figref{DecompExample}, we fill the template (i.e. by determining the correct numeric values that should appear in the code, see~\secref{fixing}), resulting in the code in block (4). The code in block (4) is then returned to the user as the final output.

For further details on the canonicalizations applied by the decompiler, see~\secref{canonization}.

\subsection{Decompilation Approach} 

Our approach to decompilation consists of two complementary phases:
\begin{inparaenum}[(1)]
\item Generating a code template that, when compiled, matches the computation structure of the input, and 
\item Filling the template with values and constants that result in code equivalent to the input.
\end{inparaenum}

\subsubsection{First Phase: Obtaining a Template} 

\figref{AcceptFigure} provides a schematic representation of this phase.

\begin{figure*}[tb]
\centering
\includegraphics[width=0.8\textwidth]{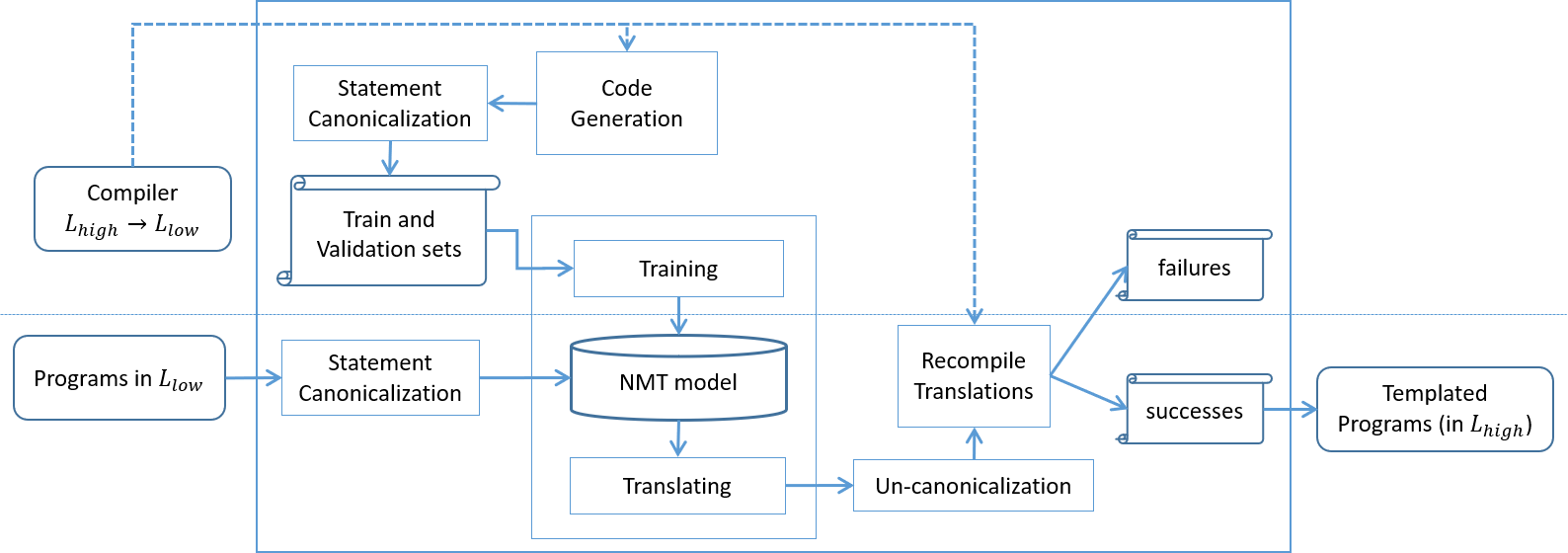}
\caption{Schematic overview of the first phase of our decompiler}
\figlabel{AcceptFigure}
\end{figure*}

At the heart of our decompiler is the NMT model. We surround the NMT model with a feedback loop that allows the system to determine success/failure rates and improve itself as needed by further training.

Denote the input language of our decompiler as $L_{low}$ and the output language as $L_{high}$, such that the grammar of both languages is known.
Given a dataset of input statements in $L_{low}$ to decompile, and a compiler from $L_{high}$ to $L_{low}$, the decompiler can either start from scratch, with an empty model, or from a previously trained model.
The decompiler translates each of the input statements to $L_{high}$. For each statement, the NMT model generates a few translations that it deemed to be most likely.
The decompiler then evaluates the generated translation. It compiles each suggested translation from $L_{high}$ to $L_{low}$ using existing of-the-shelf compilers. The compiled translations are compared against the original input statement in $L_{low}$ and classified as successful translations or failed translations. At this phase, the translations are code templates, not yet actual code, thus the comparison focuses on matching the computation structure.
A failed translation therefore does not match the structure of the input, and cannot produce code equivalent to the input in phase 2.
We denote input statements for which there was no successful translation as \emph{failed inputs}. Successful translations are passed to the second phase and made available to the user.

The existence of failed inputs triggers a retraining session. The training dataset and validation dataset (used to evaluate progress during training) are updated with additional samples, and the model resumes training using the new datasets. This feedback loop, between the failed inputs and the model's training session, drives the decompiler to improve itself and keep learning as long as it has not reached its goal. These iterations will continue until a predetermined stop condition as been met, e.g. a significant enough portion of the input statements were decompiled successfully. It also allows us to focus training on aspects where the model is weaker, as determined by the failed inputs.

The well-defined structure of programming languages allows us to make predictable and reversible modifications to both the input and output of the NMT model.
These modifications are referred to as canonicalization and un-canonicalization, and are aimed at simplifying the translation problem. These steps rely on domain specific knowledge and do not exist in traditional NMT systems for natural languages.
\secref{canonization} motivates and describes our canonicalization methods.

%Each input statement for which a successful translation was found, is removed from the input dataset and the decompiled code is made available to the user. These statements won't be translated in further iterations, thus reducing the input dataset size and the translation time in further iterations.

\para{Updating the Datasets}
After each iteration we update the dataset used for training.
Retraining without doing so would lead to over-fitting the model to the existing dataset, and will be ineffective at teaching the model to handle new inputs.

We update the dataset by adding new samples obtained from two sources:
\begin{itemize}
\item Failed translations -- We compile failed translations from $L_{high}$ to $L_{low}$ and use them as additional training samples. Training on these samples serves to teach the model the correct inputs for these translations, thus reducing the chances that the model will generate these translations again in future iterations.
\item Random samples -- we generate a predetermined number of random code samples in $L_{high}$ and compile these samples to $L_{low}$.
%These samples serve two purposes. First, the new samples guarantee there will be some advancement in the training dataset even if there is only a miniscule number of failed inputs. Second, these samples have the potential to teach the model new abilities that can prove useful to the current inputs but cannot be directly extracted from the current inputs.
\end{itemize}

The validation dataset is updated using only random samples. It is also shuffled and truncated to a constant size. The validation dataset is translated and evaluated many times during training. Thus truncating it prevents the validation overhead from increasing.

%\paragraph{Neural Machine Translation (NMT)}
%At the heart of our decompiler lies the NMT component. The goal of this component is to translate sentences from one language to another while maintaining the semantics and meaning of the original sentence. To do so, the NMT component builds a language model which it uses to parse and summarize each input sentence and for generating syntactically and grammatically correct output translations.

%The NMT component works as an encoder-decoder framework with an attention mechanism. At the input side, it summarizes the sentence and any useful context. Then, at the output side, it extracts a new sentence from the summarization. Intuitively, one can think of this framework as a compression algorithm such that the NMT component first compresses the sentence and later re-inflates it to a new language. \secref{Background} provides further technical details on NMT, the models behind it and how they works.

\subsubsection{Second Phase: Filling the Template}

The first phase of our approach produces a code template that can lead to code equivalent to the input. The goal of the second phase is to find the right values for instantiating actual code from the template. Note that the NMT model provides initial values. We need to verify that these values are correct and replace them with appropriate values if they are wrong.

This step is inspired by the common NLP practice of \emph{delexicalization}~\cite{delexicalization}. In NLP, using delexicalization, some words in a sentence would be replaced with placeholders (e.g. \scode{NAME1} instead of an actual name). After translation these placeholders would be replaced with values taken directly from the input.

Similarly, we use the input statement as the source for the values needed for filling our template. Unlike delexicalization, it is not always the case that we can take a value directly from the input. In many cases, and usually due to optimizations, we must apply some transformation to the values in the input in order to find the correct value to use.

In the example of~\figref{x86Example}, the code contains two numeric values which we need to ``fill'' -- $14$ and $2$. For each of this values we need to either verify or replace it.
The case of $14$ is relatively simple as the NMT provided a correct initial value. We can determine that by comparing $14$ in the output to $14$ in the original input.
For $2$, however, copying the value $2$ from the input did not provide the correct output. Compiling the output with the value $2$ would result in the instruction \scode{sall 1, \%eax} rather than the desired \scode{sall 2, \%eax}. We thus replace $2$ with a variable $N$ and try to find the right value for $N$. To get the correct value, we need to apply a transformation to the input. Specifically, if the input value is $x$, the relevant transformation for this example is $N = 2^x$, resulting in $N = 4$ that, when recompiled, yields the desired output. Therefore we replace $2$ with $4$, resulting in the code in~\figref{x86Example}(4).

\secref{fixing} further elaborates on this phase and provides additional possible transformations.

\section{Background}\seclabel{Background}
%\subsection{Neural Machine Translation}

Current Neural Machine Translation (NMT) models follow a sequence-to-sequence paradigm introduced in \cite{GRU}. Conceptually, they have two components, an encoder and a decoder. The encoder encodes an arbitrary length sequence of tokens $x_1,...,x_n$ over alphabet A into a sequence of vectors, where each vector represents a given input token $x_i$ in the context in which it appears. The decoder then produces an arbitrary length sequence of tokens $y_1,...,y_m$ from alphabet B, conditioned on the encoded vectors. The sequence $y_1,...,y_m$ is generated a token at a time, until generating an end-of-sequence token. When generating the $i$th token, the model considers the previously generated tokens as well as the encoded input sequence. An \emph{attention mechanism} is used to choose which subset of the encoded vectors to consider at each generation step. The generation procedure is either greedy, choosing the best continuation symbol at each step, or uses beam-search to develop several candidates in parallel. The NMT system (including the encoder, decoder and attention mechanism) is trained over many input-output sequence pairs, where the goal of the training is to produce correct output sequences for each input sequence. The encoder and the decoder are implemented as recurrent neural networks (RNNs), and in particular as specific flavors of RNNs called LSTM~\cite{LSTM} and GRU~\cite{GRU} (we use LSTMs in this work). Refer to~\cite{Neubig17} for further details on NMT systems.

\section{Decompilation with Neural Machine Translation}\seclabel{decompilation}

In this section we describe the algorithm of our decompilation framework using NMT. First, in \secref{Algo}, we describe the algorithm at a high level. We then describe the realization of operations used in the algorithm such as canonicalization (\secref{canonization}), the evaluation of the resulting translation (\secref{eval}), and the stopping condition (\secref{stop}).

\subsection{Decompiler Algorithm}\seclabel{Algo}

Our framework implements the process depicted by~\figref{AcceptFigure}. This process is also formally described in Algorithm~\ref{algorithm_alg}.
The algorithm uses a $Dataset$ data structure which holds pairs $(x,y)$ of statements such that $x\in L_{high}$, $y\in L_{low}$, and $y$ is the output of compiling $x$. 

\begin{algorithm}[tb]
  \caption{Decompilation algorithm}
	\label{algorithm_alg}
    \begin{tabular}{l l}
    \textbf{Input}  & $inputset$, collection of statements in $L_{low}$ \\
       &  $compile$, api to compile $L_{high}$ to $L_{low}$ \\
     \textbf{Output}  & Dataset of successfully decompiled\\
		& \hspace{5mm} statements in $L_{high}$ \\
     \textbf{Data Types}  & Dataset: collection of pairs $(x,y)$,\\
		& \hspace{13mm} such that $x = compile(y)$ \\
    \end{tabular}%
  \begin{algorithmic}[1]
		\Procedure{Decompile}{}
		\State $inputset \leftarrow canonicalize(inputset)$
		\State $Train \leftarrow new Dataset$
		\State $Validate \leftarrow new Dataset$
		\State $model \leftarrow new Model$
		\State $Success \leftarrow new Dataset$
		\State $Failures  \leftarrow new Dataset$
		\While {(?)}
		\State $Train \leftarrow Train \cup Failures \cup random\_samples()$
		\State $Validate = Validate \cup gen\_random\_samples()$
		\State {$model.retrain(Train, Validate)$}
		\State $decompiled \leftarrow model.translate(inputset)$
		\State $recompiled \leftarrow compile(decompiled)$
		\For{each $i$ in $0 ... inputset.size$}
		\State $pair \leftarrow (inputset[i], decompiled[i])$
		\If {$equiv(inputset[i], recompiled[i])$}
		\If {$fill(inputset[i], recompiled[i])$}
		\State $Success \leftarrow Success \cup [pair]$
		\Else
		\State $Failures \leftarrow Failures \cup [pair]$
		\EndIf
		\Else
		\State $Failures \leftarrow Failures \cup [pair]$
		\EndIf
		\EndFor
		\EndWhile
		\State \Return $uncanonicalize(Success)$
	  \EndProcedure
  \end{algorithmic}
\end{algorithm}

The framework takes two inputs:
\begin{inparaenum}[(1)]
\item a set of statements for decompilation, and
\item a compiler interface.
\end{inparaenum}
The output is a set of successfully decompiled statements.
 
Decompilation starts with empty sets for training and validation and canonicalizes~(\secref{canonization}) the input set.
It then iteratively extends the training and validation sets~(\secref{gen}), trains a model on the new sets and attempts to translate the input set.
Each translation is then recompiled and evaluated against the original input~(\secref{eval} and~\secref{fixing}). Successful translations are then put in a $Success$ set, that will eventually be returned to the user. Failed translations are put in a $Failed$ set that will be used to further extend the training set.
The framework repeats these steps as long as the stopping condition was not reached~(\secref{stop}).

\subsection{Generating Samples}\seclabel{gen}

To generate samples for our decompiler to train on, we generate random code samples from a subset of the \scode{C} programming language. This is done by sampling the grammar of the language.
The samples are guaranteed to be syntactically and grammatically correct. We then compile our code samples using the provided compiler. Doing so results in a dataset of matching pairs of statement, one in \scode{C} and the other in $L_{ll}$, that can be used by the model for training and validation.

We note that, alternatively,  we could use code snippets from publicly available code repositories as training samples, but these are less likely to cover uncommon coding patterns.

%After translation, we use the compiler again to recompile the \scode{C} code generated by the decompiler (i.e. the translations) back to the low-level language $L_{ll}$. recompilation allows us to directly compare our decompilation results with the initial input statements and verify whether or not our decompiler was successful.

\subsection{Improving Translation Performance with Canonicalization}\seclabel{canonization}

It is possible to improve the performance of NMT models without intervening in the actual model. This can be achieved by manipulating the inputs in ways that simplify the translation problem.
In the context of our work, we refer to these domain-specific manipulations as \emph{canonicalization}. 

Following are two forms of canonicalization used by our implementation:

%There are many methods for improving performance and efficiency of NMT systems, without intervening in the inner workings and models of the system. Essentially, these methods focus on simplifying the translation problem in various ways. We refer to these methods as statement canonization and wrap the NMT component with a statement canonicalization and un-canonicalization components to utilize these methods. These domain specific steps modify incoming sentences in a predictable and reversible way, such that the modified sentences would easier to process.

%While the methods depicted here are applicable to most low-level languages, some may be language specific. We thus consider this part of our technique as the only one that is not entirely language-agnostic.

\subsubsection{Reducing Vocabulary Size}\seclabel{num_abstract}
The vocabulary size of the samples provided to the model, either for training or translating, directly affects the performance and efficiency of the model.
In the case of code, a large portion of the vocabulary is devoted to numerical constants and names (such as variable names, method names, etc.).

Names and numbers are usually considered ``uncommon'' words, i.e. words that do not appear frequently. Descriptive variable names, for example, are often used within a single method but are not often reused in other methods.
%Thus, the percentage of names and numbers in code is higher than that usually found in natural languages.
This results in a distinctive vocabulary, consisting largely of uncommon words, and leading to a large vocabulary.

We observe that the actual variable names do not matter for preserving the semantics of the code. Furthermore, these names are actually removed as part of the stripping process. Therefore, we replace all names in our samples with generic names (e.g. \scode{X1} for a variable). This allows for more reuse of names in the code, and therefore more examples from which the model can learn how to treat such names. Restoring informative descriptive names in source code is a known and orthogonal research problem for which several solutions exist (e.g.~\cite{raychev2015predicting,He2018,Allamanis2015}).

Numbers cannot be handled in a similar way. Their values cannot be replaced with generic values, since that would alter the semantic meaning of the code.
Furthermore, essentially every number used in the samples becomes a word in the vocabulary. Even limiting the values of numbers to some range $[1-K]$ would still result in $K$ different words.

To deal with the abundance of numbers we take inspiration from NMT for natural languages. Whenever an NMT model for NL encounters an uncommon word, instead of trying to directly translate that word,
it falls back to a sub-word representation (i.e. process the word as several symbols).
%it revert to a separate \emph{character-level} model and process the uncommon word character by character.
Similarly, we split all numbers in our samples to digits. We train the model to handle single digits and then fuse the digits in the output into numbers. \figref{digitsSplitting} provides an example of this process on a simple input. Using this process, we reduce the portion of the vocabulary dedicated to numbers to only $10$ symbols, one per digit. Note that this reduction comes at the expense of prolonging our input sentences.

\begin{figure}[tb]
\centering
\begin{SubFloat}[]{\figlabel{digits_input}Original input}
\begin{minipage}{0.3\columnwidth}
\centering
\begin{BVerbatim}[fontsize=\small]
movl 1234 , X1
\end{BVerbatim}
\end{minipage}
\end{SubFloat}
\hfill
\begin{SubFloat}[]{\figlabel{digits_input_split}Input after splitting numbers to digits}
\begin{minipage}{0.6\columnwidth}
\centering
\begin{BVerbatim}[fontsize=\small]
movl 1 2 3 4 , X1
\end{BVerbatim}
\end{minipage}
\end{SubFloat}
\\
\begin{SubFloat}[]{\figlabel{digits_output_split}Translation output}
\begin{minipage}{0.3\columnwidth}
\centering
\begin{BVerbatim}[fontsize=\small]
X1 = 1 2 3 4 ;
\end{BVerbatim}
\end{minipage}
\end{SubFloat}
\hfill
\begin{SubFloat}[]{\figlabel{digits_output}Output after fusing digits to numbers}
\begin{minipage}{0.6\columnwidth}
\centering
\begin{BVerbatim}[fontsize=\small]
X1 = 1234 ;
\end{BVerbatim}
\end{minipage}
\end{SubFloat}
\caption{Reducing vocabulary by splitting numbers to digits}
\figlabel{digitsSplitting}
\end{figure}

\paragraph{Alternative Method for Reducing Vocabulary Size}
We observe that, in terms of usage and semantic meaning, all numbers are equivalent (other than very few specific numbers that hold special meaning, e.g. 0 and 1).
Thus, as an alternative to splitting numbers to digits, we tried replacing all numbers with constants (e.g. \scode{N1, N2, ...}). Similarly to variable names, the purpose of this replacement was to increase reuse of the relevant words while reducing the vocabulary.
When applying these replacements to our input statements, we maintained a record of all applied replacements. After translation, we used this record to restore the original values to the output.

This approach worked well for \emph{unoptimized} code, but failed on \emph{optimized} code. In unoptimized code there is a direct correlation between constants in high-level and low-level code. That correlation allowed us to restore the values in the output. In optimized code, compiler optimizations and transformations break that correlation, thus making it impossible for us to restore the output based on the kept record.

\subsubsection{Order Transformation}\seclabel{postorder}

Most high-level programming languages write code \emph{in-order}, i.e. an operator appears between its 2 operands. On the other hand, low-level programming languages, which are ''closer'' to the hardware, often use \emph{post-order}, i.e. both operands appear before the operator.

\begin{figure}[tb]
\centering
\begin{SubFloat}[]{\figlabel{postorderExampleC}original \scode{C} code}
\begin{minipage}{0.95\columnwidth}
\centering
\includegraphics[height=6mm]{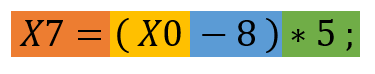}
\end{minipage}
\end{SubFloat}
\begin{SubFloat}[scale=0.5]{\figlabel{postorderExamplePostorderC}post order \scode{C} code}
\begin{minipage}{0.95\columnwidth}
\centering
\includegraphics[height=6mm]{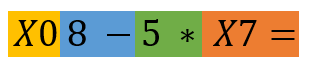}
\end{minipage}
\end{SubFloat}
\begin{SubFloat}[scale=0.5]{\figlabel{postorderExampleX86}compiled \scode{x86} assembly}
\begin{minipage}{0.95\columnwidth}
\centering
\includegraphics[height=16mm]{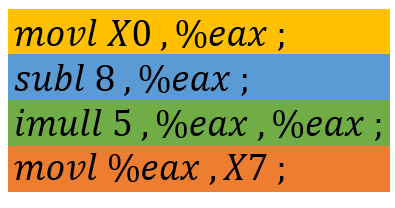}
\end{minipage}
\end{SubFloat}
\caption{Example of code structure alignment}
\figlabel{postorderExample}
\end{figure}

The code in~\figref{postorderExample} demonstrates this difference. \figref{postorderExampleC} shows a simple statement in \scode{C} and \figref{postorderExampleX86} the \scode{x86} assembly obtained by compiling it.
The different colors represent the correlation between the different parts of the computation.

Intuitively, if one was charged with the task of translating a statement, it would be helpful if both input and output shared the same order. Having a shared order simplifies ''planning'' the output by
%clustering the dependencies relevant to each part of the output.
localizing the dependencies to some area of the input rather than spreading them across the entire input.

Similarly, NMT models often perform better when when the source and target languages follow similar word orders, 
even though the model reads the entire input before generating any output.
We therefore modify the structure of the \scode{C} input statements to post-order to create a better correlation with the output.
\figref{postorderExamplePostorderC} shows the code obtained by canonicalizing the code in~\figref{postorderExampleC}.

After translation, we can easily parse the generated post-order code using a simple bottom-up parser to obtain the corresponding in-order code.
%Note that, since the decompiler now trains on code in post order, it also generates code in post order. Therefore it is required to parse the post order code back to normal code before we can recompile it. To facilitate parsing back to normal code, we added special markers that result in post order code that is easily parsable using a bottom-up parser. The decompiler trains on samples containing these markers and thus knows to generate them as part of the translations.

\subsection{Evaluating Translations}\seclabel{eval}

We rely on the deterministic nature of compilation as the basis of this evaluation.
After translating the inputs, for each pair of input $i$ and corresponding translation $t$ (i.e. the decompiled code), we recompile $t$ and compare the output to $i$. This allows us to keep track of progress and success rates, even when the correct translation is not known in advance.

\para{Comparing computation structure} 
After the first step of our decompiler, the structure of computation in the decompiled program should match the one of the original program. We therefore compare the original program and the templated program from decompilation by comparing their program dependence graphs. 
We convert each code snippet to its corresponding \emph{Program Dependence Graph} (PDG). The nodes of the graph are the different instructions in the snippet. The graph contains 2 types of edges: data dependency edges and control dependency edges.
A data dependency edge from node $n_1$ to node $n_2$ means that $n_2$ uses a value set by $n_1$. A control dependency between $n_1$ and $n_2$ means that execution of $n_2$ depends on the outcome of $n_1$. 
\figref{pdg_graph} shows an example of a program dependence graph for the code in~\figref{pdg_code}. Solid arrows in the graph represent data dependencies between code lines and dashed arrows represent control dependencies. Since line 2 uses the variable \scode{x} which was defined in line 1, we have an arrow from 1 to 2. Similarly, line 8 uses the variable \scode{z} which can be defined in either line 4 or line 6. Therefore, line 8 has a data dependency on both line 4 and line 6.
Furthermore, the execution of lines 4 and 6 is dependent on the outcome of line 3. This dependency is represented by the dashed arrows from 3 to 4 and 6.

\begin{figure}[tb]
\centering
\begin{SubFloat}[]{\figlabel{pdg_code}Source code}
\begin{minipage}{0.45\columnwidth}
\centering
  \begin{algorithmic}[1]
		\item[]
		\item[]
		\item[]
		\State $x = 3;$
		\State $y = x * x;$
		\If {$y \% 2 == 0$}
		\State $z = x + 5;$
		\Else
		\State $z = x - 7;$
		\EndIf
		\State $w = z * 2;$
		\item[]
		\item[]
  \end{algorithmic}
\end{minipage}
\end{SubFloat}
\hfill
\begin{SubFloat}[]{\figlabel{pdg_graph}Program Dependence Graph}
\begin{minipage}{0.45\columnwidth}
\centering
\begin{tikzpicture}[node distance=0.5cm, every node/.style = {shape=circle, draw, align=center}]
  \node (n1) {1};
  \node[below=of n1] (n2) {2};
  \node[below=of n2] (n3) {3};
  \node[below=of n3, color=white] (dummy) {};
  \node[right=of dummy] (n6) {6};
  \node[left=of dummy] (n4) {4};
  \node[below=of dummy] (n8) {8};
	\draw [->,>=stealth] (n1) -- (n2);
	\draw [->,>=stealth] (n1) -- (n4);
	\draw [->,>=stealth] (n1) -- (n6);
	\draw [->,>=stealth] (n2) -- (n3);
	\draw [->,>=stealth] (n4) -- (n8);
	\draw [->,>=stealth] (n6) -- (n8);
	\draw [dashed,->,>=stealth] (n3) -- (n4);
	\draw [dashed,->,>=stealth] (n3) -- (n6);
\end{tikzpicture}
\end{minipage}
\end{SubFloat}
\caption{Example of Program Dependence Graph.\\Solid arrows for data dependencies, dashed arrows for control dependencies.}
\figlabel{pdgExample}
\end{figure}
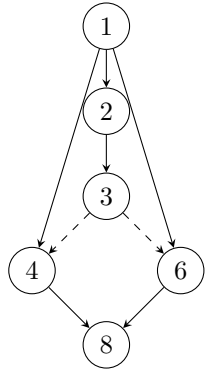

We extend the PDG with nodes ``initializing'' the different variables in the code. These nodes allow us to maintain a separation between the different variables.

We then search for an isomorphism between the 2 graphs, such that if nodes $n$ and $n'$ are matched by the isomorphism it is guaranteed that either
\begin{inparaenum}
\item both $n$ and $n'$ correspond to variables,
\item both $n$ and $n'$ correspond to numeric constants, or
\item $n$ and $n'$ correspond to the same operator (e.g. addition, substraction, branching, etc...).
\end{inparaenum}

If such an isomorphism exists, we know that both code snippets implement the same computation structure. The snippets might still differ in the variable or numeric constants they use. However, the way the snippets use these variables and constants is equivalent in both snippets.
Thus, if we could assign the correct variables and constants to the code, we would get an identical computation in both snippets. We consider translations that reach this point as a successful template and attempt to fill the template as described in~\secref{fixing}. A translation is determined fully successful only if filling the template~(\secref{fixing}) is also successful.

This kind of evaluation allows us to overcome instruction reordering, variable renaming, minor translation errors  and small modifications to the code (often due to optimizations).

\subsection{Stopping Decompilation}\seclabel{stop}

Our framework terminates the decompilation iterations when 1 of 3 conditions is met:
\begin{enumerate}
\item Sufficient results: given a percentage threshold $p$, after each iteration the framework checks the number of test samples that remain untranslated and stops when at least $p$\% of the initial test set was successfully decompiled.
\item No more progress: The framework keeps track of the amount of remaining test samples. When the framework detects that that number has not changed in $x$ iterations, meaning no progress was made during these iterations, it terminates. Such cases highlight samples that are too difficult for our decompiler to handle
\item Iteration limit: given some number $n$, we can terminate the decompilation process after $n$ iterations have finished. This criteria is optional and can be left empty, in which case only the first 2 conditions apply.
\end{enumerate}

%We note that the worst outcome would be to have experiments terminated due to \emph{no more progress}. This outcome will signify that the remaining test statements are to difficult for our framework to decompile. In practice, we haven't encountered such cases in our evaluation.

\subsection{Extending the Language}

An important feature of our framework is that we can focus the training done in the first phase to language features exhibited by the input.
Essentially, we can start by ``learning'' to decompile a subset of the high-level language.

Learning to decompile some subset $s$ of the high-level language takes time and resources. Therefore, given a new input dataset, utilizing another subset of the language $s'$, we would like to reuse what we have learned from $s$.

Because the vocabulary of $s'$ is not necessarily contained in the vocabulary of $s$, i.e. $vocab(s') \nsubseteq vocab(s)$, we have implemented a \emph{dynamic vocabulary extension mechanism} in our framework. When the framework detects that the current vocabulary is not the same as the vocabulary used for previous training sessions, it creates a new model and partially initializes it using value from a previously trained model. This allow us to add support for new tokens in the language without starting from scratch.

Note that all tokens are equivalent in the eyes of the NMT model. Specifically, the model does not know that a variable is different from a number or an operator. It only learns a difference between the tokens from the different contexts in which they appear.
Therefore, using this mechanism, we can extend the language supported by the decompiler with new operators, features and constructs, as needed.
For example, starting from a subset of the language containing only arithmetic expressions, we can easily add \scode{if} statements to the subset without losing any previous progress we've made while training on arithmetic expressions.

The extension mechanism is also used during training on a specific language subset. At each iteration, our framework generates new training samples to extend the existing training set. These new samples can, for example, contain new variables/numbers that weren't previously part of the vocabulary, thus requiring an extension of the vocabulary.

It is important to note that in a real-world use-case we don't expect training sessions to be frequent. Additional training should only applied when dealing with new features, a new language or with relatively harder samples than previous samples. We expect the majority of decompilation problems to be solved using an existing model.

\section {Filling the Template}\seclabel{fixing}

In \secref{decompilation}, we saw how the decompiler takes a low-level program and produces a high-level templated program, where some constant assignments require filling. In this section, we describe how to fill the parameters in the templated program. 

\subsection{Motivation}\seclabel{fixing_motivation}

From our experimentation with applying NMT models to code, we learned that NMT performs well at generating correct code structure. We also learned that NMT has difficulties with constants and generating/predicting the right ones. This is exhibited by many cases where the proposed translation differs from an exact translation by only a numerical constant or a variable.

The use of \emph{word embeddings} in NMT is a major contributor to these translation errors. A word embedding is essentially a summary of the different contexts in which that word appears. It is very common in NLP for identifying synonyms and other interchangeable words. For example, assume we have an NMT model for NLP which trains on the sentence ``The house is blue''. While training, the model will learn that different colors often appear in similar contexts. The model can then generalize what it has learned from ``The house is blue'' and apply that to the sentence ``The house is green'' which it has never encountered before.
In practice, word embeddings are numerical vectors, and the distance between the embeddings of words that appear in similar contexts will be smaller than the distance between embeddings of words that do not appear in similar contexts.
The model itself does not operate on the actual words provided by the user. It instead translates the input to embeddings and operates on those vectors.

Since we are dealing with code rather than natural languages, we have many more ``interchangeable'' words to handle. During training all numerical values appear in the same contexts, resulting in very similar (if not identical) embeddings. Thus, the model is often unable to distinguish between different numbers. Therefore, while word embeddings are still useful for generalizing from training examples, using embeddings in our case results in translation errors when constants are involved.

Due to the above we have decided to treat the output of the NMT model not as a final translation but as a template that needs filling. The 1st phase of our decompilation process verifies that the computation structure resulting from recompiling the translation matches that of the input. If that is the case, any differences are most likely the result of using incorrect constants.
The 2nd phase of our decompilation process deals with correcting any such false constants.

Given that the computation structure of our translation and the input is the same, errors in constants can be found in variable names and numeric values. In the first phase, as part of comparing the computation structure, we also verify that there are no cases where a variable should have been a numeric value or vice versa. That means we can treat these two cases in isolation.

We note that since we are dealing with low-level languages, in which there are often no variable names to begin with, using the correct name is inconsequential. In the case of variables, all that matter is that for each variable in the input there exists a variable in the translation that is used in exactly the same manner. This requirement is already fulfilled by matching the computation structure~(\secref{eval}).

\subsection{Finding assignments for constants}\seclabel{template_filling}
We focus on correcting errors resulting from using wrong numeric values. Denoting the input as $i$, the translation as $t$ and the result of recompiling the translation as $r$, there are three questions that we need to address:

\sectionette{Which numbers in $r$ need to change? and to which other numbers?}
Since the NMT model was trained on code containing numeric values and constants, the generated translation also contains such values (generated directly by the model) and constants (due to the numeric abstraction step we describe in~\secref{num_abstract}), and replaced with their original values. We use these numbers as an initial suggestion as to which values should be used.

As explained in~\secref{eval}, we compare $r$ and $i$ by building their corresponding program dependence graphs and looking for an isomorphism between the graphs. If such an isomorphism is found, it essentially entails a mapping from nodes in one graph to nodes in the other. Using this mapping we can search for pairs of nodes $n_r$ and $n_i$ such that $n_r \in r$ is mapped to $n_i \in i$, both nodes are numeric values, but $n_r != n_i$. Such nodes highlight which numbers need to be changed ($n_r$) and to which other numbers ($n_i$). 

\sectionette{Which numbers in $t$ affect which numbers in $r$?}
Note that although we know that $n \in r$ is wrong and to be fixed, we cannot apply the fix directly. Instead we need to apply a fix to $t$ that will result in the desired fix to $r$. The first step towards achieving that is to create a mapping from numbers in $t$ to numbers in $r$ such that changing $n_t \in t$ results in a change to $n_r \in r$.

By making small controlled changes to $t$ we can observe how $r$ is changed. We find some number $n_t \in t$, replace it with $n'_t$ resulting in $t'$ and recompile it to get $r'$. We then compare $r$ and $r'$ to verify that the change we made maintains the same low-level computation structure. If that is the case, we identify all number $n_r \in r$ that were changed and record those as affected by $n_t$.

\sectionette{How do we enact the right changes in $t$?}
At this point we know which number $n_t \in t$ we should change and we know the target value $n_i$ we want to have instead of $n_r \in r$. All we need to determine now is how to correctly modify $n_t$ to end up with $n_i$.

The simple case is such that $n_t == n_r$, which means whatever number we put in $t$ is copied directly to $r$ and thus we simply need replace $n_t$ with $n_i$.

However, due to optimizations (some applied even when using \scode{-O0}), numbers are not always copied as is.
Following are three examples we encountered in our work with \emph{x86} assembly.

\para{Replacing numbers in conditions} Assuming \scode{x} is a variable of type \scode{int}, given the code \scode{if (x >= 5)}, it is compiled to assembly equivalent to \scode{if (x > 4)}, which is semantically identical but is slightly more efficient.

\para{Division/Multiplication by powers of 2} These operations are often replaced with semantically equivalent shift operations. For example, Division by 8 would be compiled as shift right by 3.

\para{Implementing division using multiplication} Since division is usually considered the most expensive operation to execute, when the divisor is known at compilation time, it is more efficient implement the division using a sequence of multiplication and shift operations. For example, calculating $x/3$ can be done as $(x*1431655766) >> 32$ because $1431655766 \approx 2^{32}/3$.

\vspace{\baselineskip}\noindent
We identified a set of common patterns used to make such optimizations in common compilers. Using these patterns, we generate candidate replacements for $n_t$. We test each replacement by applying it to $t$, recompiling and checking whether the affected values $n_r \in r$ are now equal to their $n_i \in i$ counterparts.

We declare a translation as successful only if an appropriate fix can be found for all incorrect numeric values and constants.

\section{Evaluation}\seclabel{Eval}

In this section we describe the evaluation of our
decompilation technique and present our results.

\subsection{Implementation}\seclabel{framework}

We implemented our technique in a framework called \tool{}.
Our framework takes as input an implementation of our compiler interface and uses it to build a decompiler.
The resulting decompiler takes as input a set of sentences in a low-level language $L_{low}$, translates the sentences and outputs a corresponding set of sentences in a high-level language $L_{high}$, specifically \scode{C} in our implementation. Each sentence represents a sequence of statements in the relevant language.

Our implementation uses the NMT implementation provided by DyNmt~\cite{DyNMT} with slight modifications. DyNmt implements the standard encoder-decoder model for NMT using DyNet~\cite{dynet}, a dynamic neural network toolkit.

\paragraph{Compiler Interface}
The compiler interface consists of a set of methods encapsulating usage of the compiler and representation specific information (e.g. how does the compiler represent numbers in the assembly?).
The core of the api consists of:
\begin{inparaenum}[(1)]
\item A \scode{compile} method that takes a sequence of \scode{C} statements and returns the sequence of statements in $L_{low}$ resulting from compiling it (the returned code is ``cleaned up'' by removing parts of it that don't contribute any useful information); and
\item An \scode{Instruction} class that describes the effects of different instructions, which is used for building a PDG during translation evaluation~(\secref{eval}).
\end{inparaenum}

%The purpose of this processing step is to ''clean up'' the code, thus making it easier to decompile. For example, when compiling \scode{if} statements to \emph{LLVM IR}, the compiler generates comments that highlight the predecessors of each block. These comments might be helpful to a human researcher trying to follow the control flow of the code. However, for our decompilation needs, the comments do not add any new information and are merely making the code longer, thus making the task of translation harder.

We implemented such compiler interfaces for compilation
\begin{inparaenum}[(1)]
\item from \scode{C} to \emph{LLVM} IR, and
\item from \scode{C} to \emph{x86} assembly.
\end{inparaenum}
\figref{languagesExample} shows the result of compiling the simple \scode{C} statement of~\figref{exampleC} using both compilers.

\begin{figure}[tb]
\centering
\begin{SubFloat}[]{\figlabel{exampleC}\emph{C} code}
\begin{minipage}{0.3\columnwidth}
\centering
\begin{Verbatim}[fontsize=\small]
X0 = X1 + X2;
\end{Verbatim}
\end{minipage}
\end{SubFloat}
\\
\begin{SubFloat}[]{\figlabel{exampleLLVM}\emph{LLVM} IR}
\begin{minipage}{0.5\columnwidth}
\centering
\begin{Verbatim}[fontsize=\small]
%1 = load i32 , i32* @X1
%2 = load i32 , i32* @X2
%3 = add i32 %1 , %2
store i32 %3 , i32* @X0
\end{Verbatim}
\end{minipage}
\end{SubFloat}
\\
\begin{SubFloat}[]{\figlabel{exampleX86}\emph{x86} assembly}
\begin{minipage}{0.3\columnwidth}
\centering
\begin{Verbatim}[fontsize=\small]
movl X1 , %edx
movl X2 , %eax
addl %edx , %eax
movl %eax , X0
\end{Verbatim}
\end{minipage}
\end{SubFloat}
\caption{Example of code structure alignment}
\figlabel{languagesExample}
\end{figure}

\subsection{Benchmarks}

We evaluate \tool{} using random \scode{C} snippets sampled from a subset of the \scode{C} programming language.
Each snippet is a sequence of statements, where each statement is either an assignment of an expression to a variable, an \scode{if} condition (with or without an \scode{else} branch), or a \scode{while} loop.
Expressions consist of numbers, variables, binary operator and unary operators.
\scode{If} and \scode{while} statements are composed using a condition -- a relational operator between two expression -- and a sequence of statements which serves as the body.
We limit each sequence of statements to at most 5.
\tabref{grammartable} provides the formal grammar from which the benchmarks are sampled.
%The initial variable of the grammar is $Statements$ and terminals in the grammar are represented by an underline.

\begin{table}
	\setlength{\tabcolsep}{2pt}
  \footnotesize
  \centering
	\begin{tabular}{rcl}
	Statements&:=&Statement | Statements Statement\\
	Statement&:=&Assignment | Branch | Loop\\
	Assignments&:=&Assignment | Assignments Assignment\\
	Assignment&:=&Var = Expr;\\
	Var&:=&\underline{ID}\\
	Expr&:=&Var | \underline{Number} | BinaryExpr | UnaryExpr\\
	UnaryExpr&:=&UnaryOp Var | Var UnaryOp\\
	UnaryOp&:=&\underline{++} | \underline{--}\\
	BinaryExpr&:=&Expr BinaryOp Expr\\
	BinaryOp&:=&\underline{+} | \underline{-} | \underline{*} | \underline{/} | \underline{\%}\\
	Branch&:=&\underline{if (}Condition\underline{) \{} Statements \underline{\}} |\\
	&&\underline{if (}Condition\underline{) \{}Statements\underline{\} else \{}Statements\underline{\}}\\
	Loop&:=&\underline{while (}Condition\underline{) \{} Statements \underline{\}} |\\
	Condition&:=&Expr Relation Expr\\
	Relation&:=&\underline{>} | \underline{>=} | \underline{<} | \underline{<=} | \underline{==} | \underline{!=}
	\end{tabular}%
  \caption{Grammar for experiments. Terminals are underlined}
  \tablabel{grammartable}%
\end{table}%

All of our benchmarks were compiled using the compiler's default optimizations. Working on optimized code introduces several challenges, as mentioned in~\secref{template_filling}, but is crucial for the practicality of our approach.
Note that we didn't strip the code after compilation. However, our ''original'' \scode{C} code that we compile is already essentially stripped since our canonicalization step abstracts all names in the code. 

During benchmark generation we make sure that there is no overlap between the Training dataset, Validation dataset and our Test dataset (used as input statements to the decompiler).

\para{Evaluating Benchmarks}
Despite holding the ground-truth for our test set (the \scode{C} used to generate the set), we decided not to compare the decompiled code to the ground-truth.
We observe that, in some cases, different \scode{C} statements could be compiled to the same low-level code (e.g. the statements \scode{x = x + 1} and \scode{x++}). We decided to evaluate them in a manner that allows for such occurrences and is closer to what would be applied in a real use-case. We, thus, opted to evaluate our benchmarks by recompiling the decompiled code and comparing it against the input, as described in~\secref{eval}.

\subsection{Experimental Design and Setup}\seclabel{ExperimentalDesign}

We ran several experiments of \tool{}. For each experiment we generated 2,000 random statements to be used as the test set. \tool{} was configured to generate an initial set of 10,000 training samples and an additional 5,000 training samples at each iteration. An additional 1,000 random samples served as the validation set. There is no overlap between the test set and the training/validation sets.
We decided, at each iteration, to drop half of the training samples from the previous iteration. This serves to limit the growth of the training set (and thus the training time), and assigns a higher weight to samples obtained through recent failures compared to older samples.
Each iteration was limited to 2,000 epochs. In practice, our experiments never reached this limit. No iteration of our experiments with \emph{LLVM} and \emph{x86} exceeded more than 140 epochs (and no more than 100 epochs when excluding the first iteration). For each test input we generated 5 possible translations using \emph{beam-search}. We stopped each experiment when it has successfully translated over 95\% of the test statements or when no progress was made for the last 10 iterations.

Recall that the validation set is periodically translated during training and used to evaluate training progress. \tool{} is capable of stopping a training session early (before the epoch limit was reached) if no progress was observed in the last consecutive $k$ validation sessions.
Intuitively, this process detects when the model has reached a stable state close enough to the optimal state that can be reached on the current training set.
In our experiments a validation session is triggered after processing 1000 batches of training samples (each batch containing 32 samples) and $k$ was set to 10. All training sessions were stopped early, before reaching the epochs limit.

The NMT model consists of a single layer each for the encoder and decoder. Each layer consists of 100 nodes and the word embedding size was set to 300.
%We limited the model to inputs and outputs of length up to 2000 and 1000 respectively. In practice, none of the statements used in the test set exceeded that limit.

We ran our experiments on Amazon AWS instances. Each instance is of type \emph{r5a.2xlarge} -- a Linux machine with 8 Intel Xeon Platinum 8175M processors, each operating at 2.5GHz, and 64GiB of RAM, running Ubuntu 16.04 with GCC~\cite{gcc} version 5.4.0 and Clang~\cite{clang} version 3.8.0.

We executed our experiments as a single process using only a single CPU, without utilizing a GPU, in order to mimic the scenario of running the decompiler on an end-user's machine. This configuration highlights the applicability of our approach such that it can be used by many users without requiring specialized hardware.

\subsection{Results}\seclabel{eval_results}

\subsubsection{Estimating Problem Hardness}

As a measure of problem complexity, we first evaluated our decompiler on several different subsets of \scode{C} using only a single iteration.
The purpose of these measurements is to estimate how difficult a specific grammar is going to be for our decompiler.

We used 8 different grammars for these measurement. Each grammar is building upon the previous one, meaning that grammar $i$ contains everything in grammar $i-1$ and adds a new grammar feature (the only exception is grammar 4 which does not contain unary operators). The grammars are:
\begin{enumerate}
\item Only \emph{assignments of numbers} to variables
\item \emph{Assignments of variables} to variables
\item Computations involving \emph{unary operators}
\item Computations involving \emph{binary operators}
\item Computations involving \emph{both operators types}
\item \emph{If branches}
\item \emph{While loops}
\item \emph{Nested} branches and loops
\end{enumerate}

\begin{figure}[tb]
\centering
\includegraphics[width=0.9\columnwidth]{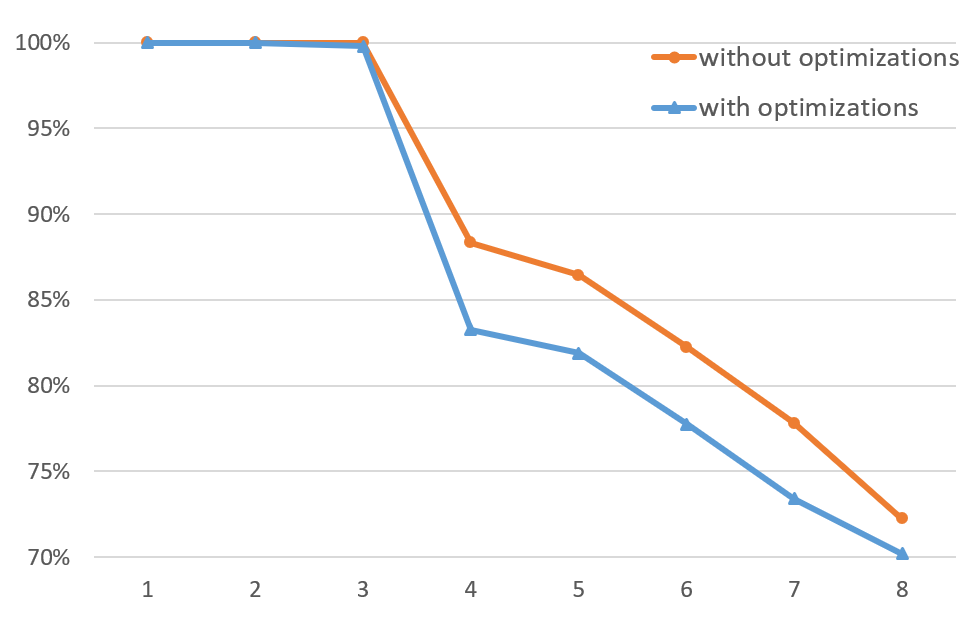}
\caption{Success rate of \emph{x86} decompiler after a single iteration on various grammars, with compiler optimization enabled and disabled}
\figlabel{single_iteration}
\end{figure}

\figref{single_iteration} shows the success rate, i.e. percentage of successfully decompiled inputs, for the different grammars, of decompiling \emph{x86} assembly with and without compiler optimizations.
Note that measured success rates are after only a single iteration of our decompilation algorithm~(\secref{Algo}).

As can be expected, the success rate drop as the complexity of the grammar increases.
That means that for more complicated grammar, our decompiler will require more iterations and/or more training data to reach the same performance level as on simpler grammars.

As can also be expected, and as can be observed from the figure, decompiling optimized code is a slightly more difficult problem for our decompiler compared to unoptimized code.
Although optimizations reduce our success rate by a few percents (at most 5\% in our experiments), it seems that the decisive factor for the hardness of the decompilation problem is the grammar complexity, not optimizations.

Recall that, given a compiler, our framework learns the inverse of that compiler.
That means that, in the eyes of the decompiler, optimizations are ``transparent''.
Optimizations only cause the decompiler to learn more complex patterns than it would have learned without optimizations, but don't increase the number of patterns learned nor the vocabulary handled.
Grammar complexity, on the other hand, increases both the number and complexity of the patterns the decompiler needs to learn and handle, and the vocabulary size, thus making the decompilation task much harder to learn.

We emphasize that enabling/disabling compiler optimizations in our framework required no changes to the framework. The only change necessary was adding the appropriate flags in the compiler interface.

\subsubsection{Iterative Decompilation}

In our second set of experiments we allowed each experiment to execute iteratively to observe the effects of multiple iterations on our decompilation success rates.

We implemented and evaluated 2 instances of our framework: from \emph{LLVM} IR to \scode{C}, and from \emph{x86} assembly to \scode{C}.

We ran each experiments $5$ times using the configuration described in~\secref{ExperimentalDesign}. We allowed each experiment to run until it reached either a success rate of$95\%$ or $6$ iterations.
The results reported below are averaged over all $5$ experiments.

\para{Decompiling \emph{LLVM} IR}
Out of the 5 experiments we conducted using our {LLVM} IR instance, 3 reached the goal of $95\%$ success rate after a single iteration. The other 2 experiments required one additional iteration to reach that goal.
\tabref{iterative_llvm} reports average statistics for these two iterations.
The columns \emph{epochs}, \emph{train time} and \emph{translate time} report averages for each iteration (i.e. average of measurements from 5 experiments for the 1st iteration and from only 2 experiments for the 2nd iteration). The \emph{successful translations} column reports the overall success rate, not just the successes in that specific iteration.

\begin{table}
  %\footnotesize
  \centering
    \begin{tabular}{c|c|c|c|c}
    & & \multicolumn{2}{c|}{timings} & successful \\
\cline{3-4} \#  & epochs & train & translate & translations \\
    \hline
    1 & 75.6 & 14:16 & 03:25 & 1913.6 (95.68\%) \\
    2 & 76.5 & 14:11 & 00:42 & 1940.2 (97.01\%) \\
    \end{tabular}%
  \caption{Statistics of iterative experiments of \emph{LLVM} IR}
  \tablabel{iterative_llvm}%
\end{table}%

The statistics in the table demonstrate that our \emph{LLVM} decompiler performed exceptionally well, even though it was decompiling optimized code snippets (which are traditionally considered harder to handle).

On average, Our \emph{LLVM} experiments successfully decompiled ~97\% of the benchmarks, before autonomously terminating. These include benchmarks consisting of up to $845$ input tokens and $286$ output tokens.
We intentionally set the goal lower than $100\%$. Setting it higher than $95\%$ and allowing our instances to run for further iterations would take longer but would also lead to a higher overall success rate.

The timing measurements reported in the table highlight that the majority of execution time is spent on training the NMT model.
Translation is very fast, taking only a few seconds per input, as witnessed by the first iteration.
The execution time of our translation evaluation (including parsing each translation into a PDG, comparing with the input PDG, and attempting to fill the templates correlating to the translations) is extremely low, taking only a couple of minutes for the entire set of benchmarks. 

%Due to the low number of failures in the 1st iteration, the training sets used in both iterations were of similar sizes, resulting in similar training times.

These observations are important due to the expected operating scenario of our decompiler. We expect the majority of inputs to be resolved using a previously trained model. Retraining an NMT model should be done only when the language grammar is extended or when significantly difficult inputs are provided.
Thus, in normal operations, the execution time of the decompiler, consisting of only translation and evaluation, will be mere seconds.

\para{Decompiling \emph{x86} Assembly}
\tabref{iterative_x86} provides statistics of our \emph{x86} experiments. 
All of these experiments terminated when they reached the iterations limit which was set to $6$.

\begin{table}
  %\footnotesize
  \centering
    \begin{tabular}{c|c|c|c|c}
    & & \multicolumn{2}{c|}{timings} & \multicolumn{1}{c}{successful} \\
\cline{3-4} \#  & \multicolumn{1}{c|}{epochs} & \multicolumn{1}{c|}{train } & \multicolumn{1}{c|}{translate} & \multicolumn{1}{c}{translations} \\
    \hline
    1 & 86.0 & 15:58 & 03:46 & 1470.8 (73.54\%) \\
    2 & 58.2 & 15:55 & 01:59 & 1614.2 (80.71\%) \\
    3 & 51.4 & 14:47 & 01:38 & 1683.2 (84.16\%) \\
    4 & 51.4 & 14:07 & 01:26 & 1721.0 (86.05\%) \\
    5 & 65.8 & 17:28 & 01:18 & 1745.6 (87.28\%) \\
    6 & 63.4 & 16:38 & 01:14 & 1762.4 (88.12\%) \\
    \end{tabular}%
  \caption{Statistics of iterative experiments of \emph{x86} asembly}
  \tablabel{iterative_x86}%
\end{table}%

\figref{cummulative_success_rate} visualizes the \emph{successful translations} column. The figure plots our average success rate as a function of the number of completed iterations.
It is evident that with each iteration the success rate increases, eventually reaching over $88\%$ after $6$ iterations.
Overall, our decompiler successfully handled samples of up $668$ input tokens and $177$ output tokens.

\begin{figure}[tb]
\centering
\includegraphics[width=0.9\columnwidth]{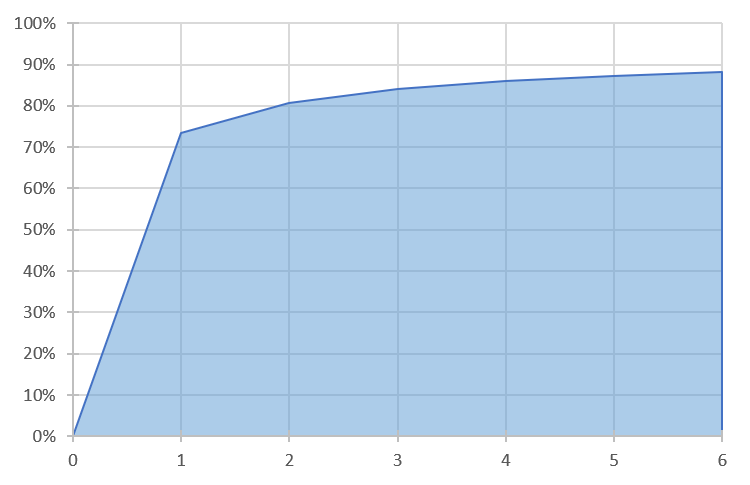}
\caption{Cummulative success rate of \emph{x86} decompiler as a function of how many iteration the decompiler performed}
\figlabel{cummulative_success_rate}
\end{figure}

%These results demonstrate our decompiler framework's capability for self-advancement when needed.
%After an initial training session, the decompiler successfully translated 78.5\% of the inputs (i.e. those that were ''easier'' for the decompiler). It then determined that these results were not sufficient and thus generated additional training examples, both random samples and samples originating from the failed translations (as explained in~\secref{Overview}), and initiated a new training session. The decompiler kept doing so until it reached the iterations limit. This entire process required no manual intervention of any kind.
%This capability is one of the main advantages of our framework over existing traditional decompilers.

Our decompilation success rates on \emph{x86} were lower than that of \emph{LLVM}, terminating at around $88\%$.
This correlates with the nature of \emph{x86} assembly, which has smaller vocabulary than that of \emph{LLVM} IR. The smaller vocabulary shortens overall training times, but also results in longer dependencies and meaningful patterns that are harder to deduce and learn.

We note that, in case of a traditional decompiler, bridging the remaining gap of $13\%$ failure rate would require a team of developers crafting additional rules and patterns. Using our technique this can be achieved by allowing the decompiler to train longer and on more training data.

\section{Discussion}

\subsection{Limitations}

Manual examination of our results from~\secref{eval_results} revealed that currently our main limitation is input length.
There was no threshold such that inputs longer than the threshold would definitely fail. We observed both successful and failed long inputs, often of the same length. We did however observe a correlation between input length and a reduced success rate.
As the length of an input increases, it becomes more likely to fail.

We found no other outstanding distinguishing features, in the code structure or used vocabulary, that we could claim are a consistent cause of failures. 

This limitation stems from the NMT model we used. long inputs are a known challenge for existing NMT systems~\cite{nmtChallenges}.
NMT for natural languages is usually limited to roughly 60 words~\cite{nmtChallenges}. Due to nature of code (i.e. limited vocabulary, limited structure) we can handle inputs much longer than typical natural language sentences (668 words for \emph{x86} and 845 words for \emph{LLVM}). Regardless, this challenge also applies to us, resulting in poorer results when handling longer inputs. As the field of NMT evolves to better handle long inputs, so would our results improve.

To verify that this limitation is not due to our specific implementation, we created another variant of our framework. This new variant is based on \emph{TensorFlow}~\cite{tensorflowSite,tensorflowArxiv} rather than DyNet.
Experimenting with this variant, we got similar results as those reported in~\secref{eval_results}, and ultimately reached the same conclusion --- the observed limitation on input length is inherent to using NMT.

\subsubsection{Other Decompilation Failures}

Though we do not consider this a limitation, another aspect that could be improved is our template filling phase. 
Our manual analysis identified some possibilities for improving our second phase -- the template filling phase~(\secref{fixing}).

\begin{figure}[tb]
\centering
\begin{SubFloat}[]{\figlabel{failure_expected}High-level code}
\begin{minipage}{0.5\columnwidth}
\centering
\begin{Verbatim}[fontsize=\small]
X3 = 63 * ( 5 * X1 ) ; 
\end{Verbatim}
\end{minipage}
\end{SubFloat}
\\
\begin{SubFloat}[]{\figlabel{failure_input}Low-level code}
\begin{minipage}{0.5\columnwidth}
\centering
\begin{Verbatim}[fontsize=\small]
movl X1 , %eax
imull 315 , %eax , %eax
movl %eax , X3
\end{Verbatim}
\end{minipage}
\end{SubFloat}
\\
\begin{SubFloat}[]{\figlabel{failure_output}Suggested decompilation}
\begin{minipage}{0.5\columnwidth}
\centering
\begin{Verbatim}[fontsize=\small]
X3 = ( X1 * 43 ) * 70 ;
\end{Verbatim}
\end{minipage}
\end{SubFloat}
\caption{Example of decompilation failure}
\figlabel{failureExample}
\end{figure}

The first type of  failure we have observed is the result of \emph{constant folding} -- a compiler optimization that replaces computations involving only constants with their results.
\figref{failureExample} demonstrates this kind of failure. Given the \scode{C} code in~\figref{failure_expected}, the compiler determines that $63 * 5$ can be replaced with $315$. Therefore, the \scode{x86} assembly in~\figref{failure_input} contains the constant $315$. Using the code of~\figref{failure_input} as input, our decompiler suggests the \scode{C} code in~\figref{failure_output}.

Note that the decompiler suggested code that is identical in structure to the input. The first phase of our decompiler handled this example correctly, resulting in a matching code template. The failure occured in the second phase, in which we were unable to find the appropriate numerical values.
This failure occurs because our current implementation attempts to find a value for each number independently from other numbers in the code. Essentially, this resulted in floating-point numbers which were deemed unacceptable by the decompiler because our benchmarks use only integers.

This kind of failure can be mitigated by either
\begin{inparaenum}[(1)]
\item applying constant folding to the high-level decompiled code,
\item allowing the template to be filled with floating point numbers (which was disabled since the benchmarks contained only integers), or
\item encoding the code as constraints and using a theorem prover to find appropriate assignments to constants.
\end{inparaenum}

%This observation emphasizes that NMT models are a suitable alternative for traditional decompilers. 

\begin{figure}[tb]
\centering
\begin{SubFloat}[]{\figlabel{failure2_expected}High-level code}
\begin{minipage}{\columnwidth}
\centering
\small
\begin{verbatim}
X2 = ((X0 \% 40) * 63) / ((98 - X1) - X0); 
\end{verbatim}
\normalsize
\end{minipage}
\end{SubFloat}
\\
\begin{SubFloat}[]{\figlabel{failure2_output}Suggested decompilation}
\begin{minipage}{\columnwidth}
\centering
\small
\begin{verbatim}
X2 = ((X0 \% N3) * N13) / (((N2 - X1) + N11) - X0);
\end{verbatim}
\normalsize
\end{minipage}
\end{SubFloat}
\caption{Failure due to redundant number}
\figlabel{failure2}
\end{figure}

A similar example is found in~\figref{failure2}. We left the suggested translation in this example as constants to simplify the example.
One can see that the suggested translation in~\figref{failure2_output} is structurally identical to the expected output in~\figref{failure2_expected}, up to the addition of $N11$.
This example was not considered a matching code template by our implementation, because any value for $N11$ other than 0 results in a different computation structure.
However, if $N11 = 0$, we get an exact match between the suggested translation and the expected output.
Using a theorem prover based template filling algorithm could detect that and assign the appropriate values to the constants, including $N11$, resulting in equivalent code.

\begin{figure}[tb]
\centering
\begin{SubFloat}[]{\figlabel{failure3_expected}High-level code}
\begin{minipage}{\columnwidth}
\centering
\small
\begin{verbatim}
X2 = 48 + (X5 * (X14 * 66));  
\end{verbatim}
\normalsize
\end{minipage}
\end{SubFloat}
\\
\begin{SubFloat}[]{\figlabel{failure3_output}Suggested decompilation}
\begin{minipage}{\columnwidth}
\centering
\small
\begin{verbatim}
X2 = ((N8 * X14) * X5) - N4; 
\end{verbatim}
\normalsize
\end{minipage}
\end{SubFloat}
\caption{Failure due to incorrect operator}
\figlabel{failure3}
\end{figure}

\figref{failure3} shows another kind of failure. In this example the difference between the expected output and suggested translation is a $+$ that was replaced with $-$.
Currently only variable names and numeric constants are treated as template parameters. This kind of difference can be overcome by considering operators as template parameters as well.
Since the number of options for each operator type (unary, binary) is extremely small, we could try all options for filling these template parameters.

\subsection{Framework Tradeoffs}

There are a few tradeoffs that should be taken into account when using our decompilation framework:
\begin{itemize}
\item Iterations limit -- Applying an iterations limit allows to tradeoff decompilation success rates for a shortened decompilation time and would make sense in environments with limited resources (time, budget, etc.). On the other hand, setting the limit too low will prevent the decompiler from reaching its full potential and will result in low successful translations rate. 
\item Training set size -- In our experiments we initialized the training set to 10,000 random samples and generated additional 5,000 new random samples each iteration. As we increase the training set size, so do the training time and memory consumption increase.
Using too many initial training samples would be wasteful in case of relatively simple test samples, in which a shorter training session, with fewer training samples, might suffice.
On the other hand, using too few samples would result in many training sessions when dealing with harder test samples. This is also applicable when setting the number of random samples added at each iteration.
Furthermore, rather than always generating a constant number of samples, one can dynamically decide the number of samples to generate based on some measure of progress (i.e. generate fewer samples when progressing at a higher rate).
\item Patience -- the patience parameter determines how many iterations to wait before terminating due to \emph{not observing any progress}. Setting this parameter to high would result in wasted time. This is because any training performed since the last time we observed progress would essentially have been in vain. On the other hand, it is possible for the model to make no progress for a few iterations only to resume progressing once it generates the training samples it needed. Setting the patience parameter too low might cause the decompiler to stop before it can reach its full potential.
\end{itemize}

\subsection{Extracting Rules}

As mentioned in~\secref{Intro}, traditional decompilers rely heavily on pattern matching. Development of such decompilers depends on hand-crafted rules and patterns, designed by experts to detect specific control-flow structures. Hand-crafting rules is slow, expensive and cumbersome.
We observe that the successful decompilations produced by our decompiler can be re-templatized to form rules that can be used by traditional decompilers, thus simplifying traditional decompiler development.
\appref{rules} provides examples of such rules.

\subsection{Evaluating Readability}

Measuring the readability of our translations requires a user study, which we did not perform.
However, note that given some training set, a model trained on that set will generate code that is similar to what it was trained on. Thus, the readability of our translations stems from the readability of our training samples. Our translations are as readable as the training samples we generated. This was also verified by an empirical sampling of our results.
Therefore, given readable code as training samples, we can surmise that any decompiled code we generate and output will also be readable.

%!TEX root=./main.tex

\section{Related Work}\seclabel{Related}

\para{Decompilation}
The Hex-Rays decompiler~\cite{IDA} was considered the state of the art in decompilation, and is still considered the \emph{de-facto} industry standard.
Schwartz et al.~\cite{Phoenix2013} presented the Phoenix decompiler which improved upon Hex-Rays using new analysis techniques and iterative refinement, but was still unable to guarantee \scode{goto}-free code (since \scode{goto} instructions are rarely used in practice, they should not be part of the decompiler output).
Yakdan et al.~\cite{Dream2015,Dream++2016} introduced Dream, and its predecessor Dream++, taking a significant step forward by guaranteeing \scode{goto}-free code. 
RetDec~\cite{retdec}, short for \emph{Retargetable Decompiler}, is an open-source decompiler released in December 2017 by Avast, aiming to be the first ''generic'' decompiler capable of supporting many architectures, languages, ABIs, etc.

While previous work made significant improvements to decompilation, all previous work fall under the title of rule-based decompilers. Rule-based decompilers require manually written rules and patterns to detect known control-flow structures. These rules are very hard to develop, prone to errors and usually only capture part of the known control-flow structures.
According to data published by Avast, it took a team of ~24 developers ~7 years to develop RetDec.
This data emphasizes that traditional decompiler development is extremely difficult and time consuming, supporting our claim that the future of decompilers lies in approaches that can avoid this step.
Our technique removes the burden of rule writing from the developer, replacing it with an automatic, neural network based approach that can autonomously extract relevant patterns from the data.

Katz et al.~\cite{saner2018} suggested the first technique to use NMT for decompilation.
While they set out to solve the same problem, in practice they provide a solution to a different and significantly easier problem - producing source-level code that is readable, without any guarantees for equivalence, not semantic or even syntactic. Further, the code they generate is not guaranteed to compile (and does not in practice). Because their code does not compile nor is equivalent, if you apply our evaluation criteria to their results, their accuracy would be at most 3.8\%. 
Further, beyond the cardinal difference in the problem itself, they have the following limitations:
\begin{compactitem}
\item They can only operate on code compiled with a special version of Clang which they modified for their purposes.
\item All of their benchmarks are compiled without optimizations. We apply the compiler's default optimizations to all of our benchmarks.
\item They limit their input to 112 tokens and output to 88 tokens. This limits their input to single statements. We successfully decompiled \emph{x86} benchmarks of up to $668$ input tokens and $177$ output tokens. Each of our samples contains several statements.
\item Their methodology is flawed as they do not control for overlaps between the training and test sets. We verify that there is no such overlap in our sets.
\end{compactitem}

\para{Modeling Source Code}
Modeling source code using various statistical models has seen a lot of interest for various applications.

Srinivasan et al.~\cite{Srinivasan2016} used LSTMs to generate natural language descriptions for C\# source code snippets and SQL queries.
Allamanis et al.~\cite{Allamanis2016} generated descriptions for Java source code using convolutional neural networks with attention. Hu et al.~\cite{Hu2017}  tackled the same problem by neural networks with a structured based traversal of the abstract syntax tree, aimed at better representing the structure of the code. 
Loyola et al.~\cite{Loyola2017} took a similar approach for generating descriptions of changes in source code, i.e. translates commits to source code repositories to commit messages.
The success presented by these papers highlights that neural networks are useful for summarizing code, and supports the use of neural networks for decompilation.

Another application of source code modeling is for predicting names for variable, methods and classes. 
Raychev et al.~\cite{Raychev2015} used conditional random fields (CRFs) to predict variable names in obfuscated JavaScript code. He et al.~\cite{He2018} also used CRFs but for the purpose of predicting debug information in stripped binaries, focusing on names and types of variables.
Allamanis et al.~\cite{Allamanis2015B} used neural language models to predict variable, method and class names. Allamanis et al. relied on word embeddings to determine semantically similar names.
We consider this problem as orthogonal to our own. Given a semantically equivalent source code produced by our decompiler, these techniques could be used to supplement it with variable names, etc.

Chen et al.~\cite{Chen2018} used neural networks to translate code between high-level programming languages. This problem resembles that of decompilation, but is infact simpler. Translating low-level languages to high-level languages, as we do, is more challenging. The similarities between high-level languages are more prevalent than between high-level and low-level languages. Furthermore, translating source code to source code directly bypasses many challenges added by compilation and optimizations.

Levy et al.~\cite{Levy2017} used neural networks to predict alignment between source code and compiled object code. Their results can be useful in improving our second phase, i.e. filling the template and correcting errors. Specifically, their alignment prediction can be utilized to pinpoint location in the source code that lead to errors.

Katz et al.~\cite{myPopl2016, myAsplos2018} used statistical language models for modeling of binary code and aspects of program structure. Based on a combination of static analysis and simple statistical language models they predict targets of virtual function calls~\cite{myPopl2016} and inheritance relations between types~\cite{myAsplos2018}. Their work further highlights that these techniques can deduce high-level information from low-level representation in binaries.

%Maddison et al.~\cite{Maddison2014} and Amodio et al.~\cite{Amodio2017} have both used statistical models to generate code. Maddison et al. used while probabilistic context free grammars and the more recent Amodio et al. used neural attribute machine, a form of RNN that also utilizes language grammar. The code generation problem resembles the problem we deal with, in that the outcome must be syntactically and grammatically valid.
%!TEX root=./main.tex

\section{Conclusion}\seclabel{Conclusion}

We address the problem of \emph{decompilation} --- converting low-level code to high-level human-readable source code.
Decompilation is extremely useful to security researchers as the cost of finding vulnerabilities and understanding malware drastically drops when source code is available.
%It also as other applications such as porting software to new architectures/operating systems.

A major problem of traditional decompilers is that they are rule-based. This means that experts are needed for hand-crafting the rules and patterns used for detecting control-flow structures and idioms in low-level code and lift them to source level. As a result decompiler development is very costly.

We presented a new approach to the decompilation problem. We base our decompiler framework on \emph{neural machine translation}. Given a compiler, our framework automatically learns a decompiler from it. We implemented an instance of our framework for decompiling \emph{LLVM} IR and \emph{x86} assembly to \scode{C}. We evaluated these instances on randomly generated inputs with a high success rates.

%%%%%%%%%%%%%%%%%%%%%%%%%%%%%%%%%%%%%%%%%%%%%

%% Bibliography style
\bibliographystyle{ACM-Reference-Format}
\bibliography{bib}

%\clearpage
\appendix
%\clearpage

\clearpage
\begin{landscape}

\section{Extracting Decompilation Rules}\applabel{rules}

\tabref{translation_rules} contains examples of decompilation rules extracted from our decompiler. For brevity, we present mostly relatively simple rules, but longer and more complicated rules were also found by our decompiler (examples of such rules are found at the bottom of the table, below the separating line).

\begin{center}
\begin{table}[hp!]
  \footnotesize
    \begin{tabular}{l|l}
input  & output \\
    \hline
movl $X_1$ , eax ; addl $N_1$ , eax ; movl eax , $X_2$ ; & $X_2 = N_1 + X_1 ;$ \\
movl $X_1$ , eax ; subl $N_1$ , eax ; movl eax , $X_2$ ; & $X_2 = X0 - N_1 ;$ \\
movl $X_1$ , eax ; imull $N_1$ , eax , eax ; movl eax , $X_2$ ; & $X_2 = X_1 * N_1 ;$ \\
movl $X_1$ , ecx ; movl $N_1$ , eax ; idivl ecx ; movl eax , $X_2$ ; & $X_2 = N_1 / X_1 ;$ \\
movl $X_1$ , eax ; movl $X_2$ , ecx ; idivl ecx ; movl eax , $X_3$ ; & $X_3 = X_1 / X_2 ;$ \\
movl $X_1$ , eax ; sall $N_1$ , eax ; movl eax , $X_2$ ; & $X_2 = X_1 * 2 ^ {N_1} ;$ \\
movl $X_1$ , ecx ; movl $N_1$ , eax ; idivl ecx ; movl edx , eax ; movl eax , $X_2$ ; & $X_2 = N_1 \% X_1 ;$ \\
movl $X_1$ , eax ; movl $X_2$ , ecx ; idivl ecx ; movl edx , eax ; movl eax , $X_3$ ; & $X_3 = X_1 \% X_2 ;$ \\
movl $X_1$ , eax ; leal 1 ( eax ) , edx ; movl edx , $X_1$ ; movl eax , $X_2$ ; & $X_2 = X_1\texttt{+}\texttt{+} ;$ \\
movl $X_1$ , eax ; leal -1 ( eax ) , edx ; movl edx , $X_1$ ; movl eax , $X_2$ ; & $X_2 = X_1\texttt{-}\texttt{-} ;$ \\
movl $X_1$ , eax ; addl 1 , eax ; movl eax , $X_1$ ; movl $X_1$ , eax ; movl eax , $X_2$ ; & $X_2 = \texttt{+}\texttt{+}X_1 ;$ \\
%movl $X_1$ , eax ; subl 1 , eax ; movl eax , $X_1$ ; movl $X_1$ , eax ; movl eax , $X_2$ ; & $X_2 = \texttt{-}\texttt{-}X_1 ;$ \\
movl $X_1$ , eax ; imull $N_1$ , eax , eax ; addl $N_2$ , eax ; movl eax , $X_2$ ; & $X_2 = N_2 + (N_1 * X_1) ;$ \\
movl $X_1$ , eax ; addl $N_1$ , eax ; sall $N_2$ , eax ; movl eax , $X_2$ ; & $X_2 = (X_1 + N_1) * 2 ^ {N_2} ;$ \\
movl $X_1$ , eax ; imull $N_1$ , eax , ecx ; movl $N_2$ , eax ; idivl ecx ; movl eax , $X_2$ ; & $X_2 = N_2 / ( X_1 * N_1 ) ;$ \\
movl $X_1$ , eax ; cmpl $N_2$ , eax ; jg .L0 ; movl $N_2$ , $X_2$ ; .L0: ; & $\emph{if} (X_1 < (N_1 + 1)) \{ X_2 = N_2 ; \}$ \\
jmp .L1 ; .L0: ; movl $N_1$ , $X_1$ ; .L1: ; movl $X_2$ , eax ; cmpl $N_2$ , eax ; jg .L0 ; & $\emph{while} (X_2 > N_2) \{ X_1 = N_1 ; \}$ \\
jmp .L1 ; .L0: ; movl $N_1$ , $X_1$ ; .L1: ; movl $X_2$ , eax ; cmpl $N_2$ , eax ; jne .L0 ; & $\emph{while} (N_2 != X ) \{ X_1 = N_1 ; \}$ \\
movl $X_1$ , eax ; cmpl $N_1$ , eax ; jne .L0 ; movl $N_2$ , $X_2$ ; movl $X_3$ , eax ; movl eax , $X_4$ ; .L0: ; & $\emph{if} (N_1 == X_1) \{ X_2 = N_2 ; X_4 = X_3 ; \}$ \\
movl $X_1$ , edx ; movl $X_2$ , eax ; cmpl eax , edx ; jg .L0 ; movl $N_1$ , $X_3$ ; jmp .L1 ; .L0: ; movl $N_2$ , $X_4$ ; .L1: ; & $\emph{if} (X_1 <= X_2) \{ X_3 = N_1 ; \} \emph{else} \{ X_4 = N_2 ; \}$ \\
jmp .L1 ; .L0: ; movl $X_1$ , eax ; addl $N_1$ , eax ; movl eax , $X_2$ ; .L1: ; movl $X_3$ , eax ; cmpl $N_2$ , eax ; jle .L0 ; & $\emph{while} (X_3 <= N_2) \{ X_2 = N_1 + X_1 ; \}$ \\
\hline
jmp .L1 ; .L0 : ; movl $X_2$ , eax ; addl 1 , eax ; movl eax , $X_2$ ; movl $X_2$ , edx ; movl $X_2$ , eax ; addl edx , ea... & $\emph{while} ((X_1 - N_1) > (X_2 \% (X_2 - N_2))) \{ X_3 = (\texttt{+}\texttt{+}X_2) + X_2 ; ...$ \\
movl $X_1$ , eax ; addl 1 , eax ; movl eax , $X_1$ ; movl $X_1$ , edx ; movl $X_2$ , eax ; movl $N_1$ , ecx ; subl eax , ecx... & $\emph{if} ( \texttt{+}\texttt{+} X_1 == ( ( ( X_2 * ( N_1 - X_2 ) ) - N_2 ) * ( N_3 - X_3 ) ) ) \{ X_2 =...$ \\
movl $X_3$  , edx ; movl $X_4$  , eax ; addl edx , eax ; movl $X_4$  , ecx ; movl $X_5$ , edx ; addl edx , ecx ; idivl ecx ; ... & $X_1  = X_2  * ( ( X_3  + X_4  ) \% ( X_4  + X_5 ) ) ; X_6 = ( X_7 + X_9 ) / ( ( N_1 - ...$ \\
movl $N_1$ , $X_1$ ; movl $X_1$ , eax ; movl eax , $X_2$ ; movl $X_2$ , eax ; movl $X_3$ , edx ; addl $N_3$ , edx ; subl edx , ea... & $X_1 = N_1 ; X_2 = X_1 ; if ( ( N_2 + ( X_2 - ( X_3 + N_3 ) ) ) <= X_4 ) \{ X...$ \\
jmp .L1 ; .L0 : ; movl $X_1$ , ebx ; movl $N_3$ , eax ; idivl ebx ; movl eax , $X_1$ ; .L1 : ; movl $X_1$ , edx ; movl $X_2$ ... & $while ( ( X_1 * X_2 ) >= ( N_1 \% ( X_3 + N_2 ) ) ) \{ X_1 = N_3 / X_1 ; \} ; X_4 ...$ \\
    \end{tabular}
\caption{Decompilation rules extracted from \tool{}}
\tablabel{translation_rules}
\end{table}
\end{center}

\end{landscape}

\end{document}